\documentclass[12pt]{article}  
   \usepackage{doublespace}  
\usepackage{epsfig} 
\usepackage{graphics} 
\usepackage{rotate} 

\def \bg{\bigskip} 
\def \no{\noindent} 
\begin{document}

%\normalsize
  \small

\setcounter{page}{1}
%\bg 

%  GRG   

%  accel3,,, ,  
% Nov. 20  1999  
% .

\bg

%\bg

\bg

\begin{center} 
{\large {\bf THE ORIGIN OF THE RADIATION REACTION FORCE }}  

\end{center}
%\bg

%\bg
\bg
{\centerline  {\bf Amos Harpaz  \& Noam Soker }} 

\bg

{\centerline { Department of Physics, University of Haifa at 
Oranim, Tivon 36006, ISRAEL }}
   
 \no  phr89ah@tx.technion.ac.il 

 \no soker@physics.technion.ac.il

\bg
\no {\bf ABSTRACT} 
\bg

 The emission of radiation from an accelerated charge is analyzed. 
 It is found that at zero velocity, the radiation emitted from 
the charge imparts no counter momentum to the emitting charge, 
and no radiation reaction force is created by the radiation.  A 
reaction force is created by the stress force that exists 
in the curved electric field of the charge, and the work done in overcoming 
this force is the source of the energy carried by 
the radiation.  

\bg

\no  PACS: 04.,95.30

\no key words: Radiation Reaction Force, Curved Electric Field 

\vfil

\eject

\bg

\no {\bf 1.  Introduction} 
\bg

        A radiation emitted by an accelerated charge carries energy 
  generated through the process of the creation of the radiation.  
Comparing accelerations of neutral and charged particles having  
equal masses, we find that for both particles, the work done by the 
accelerating  force creates the kinetic energy of the particles.  However, 
the acceleration of the charged particle generates additional energy, 
the energy carried by the radiation, and this additional energy is 
generated by an additional work done by the accelerating force.  Thus 
the accelerating force of a charged particle may be decomposed into 
two parts: one, is the force  that works against inertia, and it generates 
the kinetic energy of the particle.  The second part, the one that 
generates the energy carried by the radiation, works against a 
reaction force, which is usually called ``Radiation reaction force" 
(ref. [1],[2],[3]).   It is assumed that when 
radiation is emitted   
 by an accelerated charge, a reaction force is created, which contradicts 
the acceleration, and  the work done in overcoming this force, is 
the source of the energy carried by the radiation.  

 However, it comes out that not all the radiated power creates a reaction 
force, and when the velocity of the accelerated charge is very low, this 
reaction force actually vanishes.  A question is raised, what is then the 
source of the energy carried by the radiation.  
We find that, when a charged  particle is accelerated in a free space, 
its electric field, which is an independent physical entity, is not 
accelerated with the charge.   The electric field is curved [4], 
 and a stress force exists between the accelerated charge and its curved 
 electric field [5].  This force is (a part of) the reaction force, 
 and the work done in overcoming this force is the source of the 
energy carried by the radiation.  If the velocity of the accelerated charge 
is not zero, the radiation creates part of the 
 reaction force, which together with the stress force of the curved 
electric field, forms the total radiation reaction force.

In \S 2 we present the problem that arises when the {\it radiation  }   
reaction force vanishes.   
   In \S 3 we present the  solution to the problem,  where 
 a reaction force is shown to exist due to the stress force in the 
 curved electric field of the charge, even at zero velocity.  
  In \S 4  we calculate the reaction force for non-zero velocity, 
  and we conclude in \S 5.    

\bg
  
\no {\bf 2. The Problem } 
\bg

 The  formula for the angular distribution  
of the radiation power is [6]: 

$$ {dP\over d\Omega} = {e^2 a^2\over 4\pi c^3} {\sin^2\theta 
\over (1 - \beta \cos\theta)^5}. \eqno(1)  $$ 

\no where $e$ is the accelerated  charge,  $a$ is its acceleration,  
 $\theta $ is the angle measured from the direction of the acceleration,  
  and $\beta$ is the velocity of the particle divided by the speed 
 of light ($c$).     

Integrating eq. 1 over the angles, yields:
 $$P = {2\over 3}{e^2 (\gamma^3  a)^2\over c^3 }  \eqno(2) $$

\no where $\gamma^2 = 1/(1-\beta^2)$.  Equation 2 
 yields Larmor formula for the power carried by radiation 
 for zero velocity  ($\beta \rightarrow 0$):  
 
$$  P = {2\over 3}{e^2 a^2\over c^3} .  \eqno(3)  $$

We plot in Fig. 1 the angular  distribution of the radiation 
 according to eq. 1, for 
several values of $\beta$  (similar figure is given in [6], 
 (his Fig. 14.3)).  

\begin{figure}
\centering\epsfig{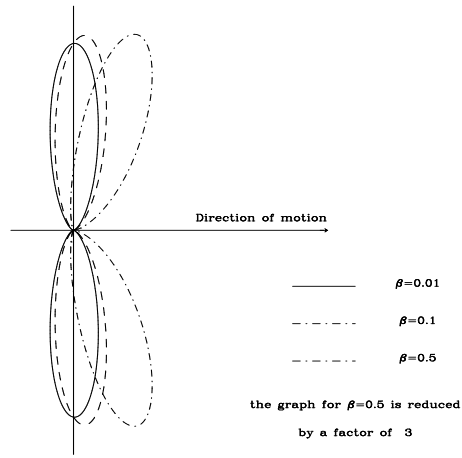} 
\vskip 0.5cm
\caption{The angular distribution of the radiation.} 
\end{figure} 

It is clearly observed that for low velocities $(\beta \le 0.01)$, 
 most of the radiation is emitted  at a right angle to the 
direction of motion, and due to the symmetry in the plane 
perpendicular to the direction of motion, no counter momentum is 
imparted to the emitting particle, and actually, no reaction force 
is created by the radiation. 
  A trivial question is raised - what is the source of the 
 energy carried by the radiation.  This problem bothered scientists 
 (it was named as the
 ``Energy Balance Paradox"), and several answers were suggested. 

One of them was [7], that there exists a charged plane, 
 whose charge is equal and opposite in sign to the accelerated 
 charge, and it recesses with the speed of light in a direction 
opposite  to the direction of the acceleration.  The interaction between 
 this charged plane and the accelerated charge creates the energy 
carried by the radiation.  
Another suggestion was [2],  that in such a case the 
energy radiated is supplied from the self-energy of the charge.  
Evidently, these suggestions are far from being satisfactory. 
 It should be noted that the idea suggested in  [7]  resolves another 
difficulty concerning this topic, which is the existence of a single 
electric charge.   As we assume that the matter in the universe is 
neutral, the existence of a solitary charge is a local phenomenon, whose 
validity is limited to distance scales that are much shorter than 
distance scales that characterize gravitational considerations.  Any 
treatment of this topic that carries calculations to infinity, cannot be a 
valid treatment.  The treatment suggested by Leibovitz and Peres [7], 
considers a system which is neutral.   

The most simple case of acceleration is the one in which the particle 
 is accelerated with a constant acceleration, $a$, in its own system 
 of reference.  In  such a motion, the particle is always at rest in a  
 system of reference which is adjacent to the particle.   Such a motion is 
 characterized  as a hyperbolic motion [8].  In such a motion, 
 the radiation 
 reaction force calculated from Dirac equation of motion vanishes  
 for zero velocity, exactly as we observe it in Fig. 1. It becomes  
 evident that this reaction force cannot create the energy carried 
 by the radiation, and the source of this energy should be looked 
 for elsewhere.  

\bg

\no {\bf 3. The Solution } 
\bg

An electric field is an independent physical entity, whose 
behaviour depends on the properties of space on which it is induced.  
 The electric field is induced by an electric charge, but once it 
is induced, its properties do not depend any more on the charge 
that has induced it.  The electric field is inertial with respect 
to the local (free) system of reference, characterized by the 
geodesics that cover this system.  
 Thus, when an accelerating charge induces 
an electric field, the field {\it is not accelerated} with the charge, 
 and there exists a relative acceleration between the field and the 
charge that induced the field. (The existence of an acceleration 
between the charge and its electric field is the condition for the 
creation of radiation - and not the relative acceleration between the 
charge and the observer, as was suggested by several authors (see [5])).  
 Let us consider the case of a constant acceleration, characterized as 
 a hyperbolic motion. 

\begin{figure}
\centering\epsfig{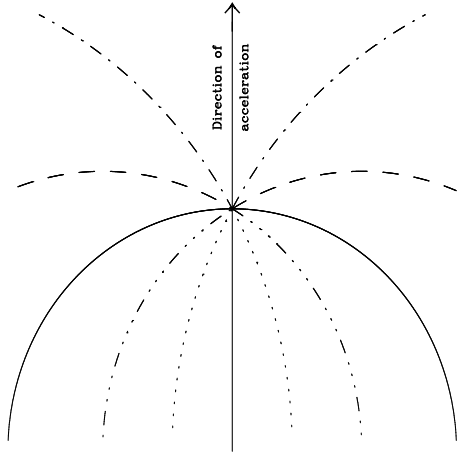} 
\vskip 0.5cm
\caption{A curved electric field of a uniformly accelerated charge.} 
\end{figure}

 The field induced by the accelerating charge is 
curved, and this curvature is displayed in Fig. 2 (this graph is 
similar to the one given by Singal [9])  Due to the curvature there 
exists a stress force in the field, whose force density is given by: 

$$ f_E = {E^2\over 4 \pi R_c}     \eqno(4) $$ 

\no where  $E$ is the field strength, and  $R_c$  is the radius of curvature 
of the field.  The interaction between the accelerated charge and its 
curved electric  field creates a force that contradicts the  acceleration.  
In order to overcome this force, the accelerating force should perform 
 additional work (in addition to the work that creates the kinetic energy 
of the particle), and this additional work  creates the energy  carried 
by the radiation.  We have to sum over the stress force, $f_E$, and   
calculate the work done against this force.

In order to sum over $f_E$,   we have to 
integrate over a sphere whose center is located on the charge.  
Naturally, such an integration involves a divergence (at the center). 
To avoid such a divergence, we take as the lower limit of the 
integration a small distance from the center, $r = c \Delta t$, (where 
$\Delta t$ is infinitsimal).  
  We calculate the work done by the stress 
force in the volume defined by  $c \Delta t < r < r_{up}$, 
where  $c^2/a \gg r_{up} \gg c \Delta t$.   
These calculations are performed in a system of reference $S$, 
defined by the geodesics, and momentarily coincides with the frame of 
reference of the accelerated charge  
at time $t = 0$, at the charge location (see ref. [5]).   

 The force per unit volume  due to the electric stress is 
$f_E = E^2/(4 \pi R_c)$.  
 The radius of curvature is calculated by using the formulae for 
  the field lines [9].   
 It can be easily shown that in the limit of $a \Delta t \ll c$,   
the radius of curvature of a field line 
 is $R_c \simeq  c^2 / (a \sin \theta)$,
where $\theta$ is the angle between the initial direction of the 
 field line 
and the acceleration (see Fig. 2).   
 The force per unit volume due  to   the electric stress is

$$   f_E (r) = {{E^2}(r)\over{4 \pi R_c}} = 
 {{a \sin \theta}\over{c^2}} {{e^2}\over{4 \pi r^4}} , \eqno(5)  $$   

\no where in the second equality we have substituted for the electric 
field $E=e/r^2$,  which is a good approximation for low acceleration  [4].  
 The stress force is perpendicular to the direction of the field lines, 
so that the component of the stress force along the acceleration is
$ - f_E(r) \sin \phi$, where  $\phi$ is the angle between the 
local field line and the acceleration.   For very short intervals 
 ($\ll c^2/a$,  where 
the direction of the  field lines did not change much 
 from their original  direction) $\phi \sim \theta$, 
and we can write: 
$ - f_E(r) \sin \phi \simeq - f_E(r) \sin \theta  = 
  {-a \sin^2 \theta \over c^2} {e^2 \over 4 \pi r^4}$.   
The dependence of this force on $\theta$ is similar to the dependence of the radiation 
distribution of an  accelerated charge at zero velocity on $\theta$.      
 Since the calculations are performed in the system  $S$, which is 
a flat space system (defined by the geodesics), the integration 
can be carried without using any terms concerning space curvature.  
Integration of the stress force  over a spherical shell 
 extending from $r=c \Delta t$ 
to $r_ {\rm up}$, where $c^2 / a \gg r_{\rm up} \gg c \Delta t$,
yields the total force due to stress 

$$  F_E(t) = 2 \pi  \int _{c \Delta t}^{r_{\rm up}} r^2 dr 
\int_0^{\pi} \sin \theta d \theta
[- f_E(r) \sin \theta ]
= - {{2}\over{3}} {{a}\over{c^2}} {{e^2}\over{c \Delta t}}
\left(1-{{c \Delta t}\over{r_{\rm up}}} \right) . \eqno(6)  $$
 
 Clearly the second term in the parenthesis can be neglected.
The power supplied by the external force on acting against the
electric stress is $P_E= - F_E  v = - F_E a \Delta t $, 
where we substituted  $v= a \Delta t$, and  $v$  is the charge velocity 
 at time $t = \Delta t$, in the system $S$, the system defined by the 
geodesics, and mometarily coincides with the system of reference of 
the accelerated charge st time $t = 0$, at the charge location.        
Substituting for  $F_E$   we obtain 

$$ P_E(t) = {{2}\over{3}} {{a^2 e^2}\over{c^3}} . \eqno(7)   $$  

 This is the power radiated by an accelerated charged particle 
 at zero velocity (eq. 3). 

Actually, the expression found in eq. 6 before substituting the limits 
of the integral ($F_{(E)} = -{2\over 3} ae^2/c^2r$), equals  the inertial 
force ($4 m_ea/3$) of the electromagnetic mass of the charge, as 
 calculated by Lorentz (see ref [6] pp 790). However, the force calculated, 
 here is entirely a different one.   A work done against 
 any inertial force, creates 
kinetic energy of the relevant mass, and cannot be a source of the energy 
carried by the radiation.  This leaves us with the energy balance paradox.  
The work done in overcoming the stress force (calculated in eq. 6) is 
done in addition to any work done against inertial forces that create 
kinetic anergy, and this work is the source of the 
energy carried by the radiation.    

\bg
\no{\bf  4. Non-Zero Velocity  }
\bg

 Equation (7) shows the power emitted by a uniformly 
accelerated charge calculated at zero velocity.  We can consider   
 the case of an accelerated charge, moving with low velocity. 
 From Fig. 1 we observe that part of the radiation is radiated forward, 
 and evidently, this part of the radiation imparts a backward momentum 
to the radiating charge, thus creating a  reaction force.   
 To calculate the power created by {\it this} reaction force, we  
 multiply eq. 1  by $\cos \theta/c$, (to obtain the parallel 
component of the momentum flux of the 
 radiation), and integrate over the angles.  The 
integration yields for the parallel component of the
momentum flux   $(p_{par})$:  

$$p_{par} = {2\over 3}{e^2 (\gamma^3a)^2\over c^4} \beta , \eqno(8) $$   

\no  while the total absolute value of the momentum flux,  
 $p$,  is found  by dividing  eq. 2 by $c$.  
(The perpendicular momentum flux vanishes because of the 
symmetry mentioned above, but we still can compare the parallel 
component of the momentum flux to the total absolute value of the 
momentum flux of the radiation).   By dividing $p_{par}$ by $p$,  
we find the weight of the parallel component of the momentum flux in 
the total absolute value of the momentum flux:

$${p_{par}\over p} =  \beta     \eqno(9)  $$

This fraction is the weight of the parallel component of the 
momentum flux of the radiation, and this fraction creates a reaction
force - a radiation reaction force. 
 To get the weight of the work done by the radiation reaction 
force we should multiply this fraction by $\beta c$ (the velocity 
of the charge in the rest frame), and we find that 
 the weight of the work 
done in overcoming  the radiation reaction 
force (the reaction force created by the radiation) 
 in the total power radiated is  $\beta^2$.   The other part of the 
energy $( 1 - \beta^2 = 1/\gamma^2) $, is created by the stress force that 
 exists in the curved electric field.    
We find that the weight of the work done by the stress force in the total 
work done by the total reaction force, decreases when $\gamma$ increases.  

\bg
\no{\bf  5.  Conclusions }
\bg

It is shown that for the case of an acceleration of a charge at zero 
 velocity, the emission of radiation by the accelerated charge does 
not create any reaction force, exactly as it is found from Dirac's 
 equation of motion for the electron.   In such a case, the reaction force 
is the stress force created by the curved electric field, which does not 
 participate in the acceleration of the charge.  When the velocity of the 
accelerated charge is not zero, part of the radiation is emitted forward, 
 and contributes to the reaction force, while another part of the reaction 
force is contributed by the stress force in the curved electric field.  
 The total reaction force is the sum of the contributions of these two parts.  

   \bg

 \bg

 \vfil
  \eject 
                                                         
\no {\bf references}
\bg

 \no [1] Wheeler, J.A., Feinman, R.P., 1945, Rev. Mod. Physics, 
 17, 157.

\no [2] Fulton, A., Rohrlich, F., 1960, Annals of Physics, 9, 499.   

\no [3] Teitlboim, C., 1970, Physical Rev. D, 1, 1572.  

 \no [4]  Rohrlich, F. 1965, in {\it Classical Charged Particles}, 
  Addison-Wesley Pub. Co., MA.   

 \no [5] Harpaz, A., Soker, N., 1998,  
    Gen. Rel. Grav., 30, 1217.

 \no [6]  Jackson, J. D. 1975, {\it Classical Electrodynamics}, 
Second Edition, John Wiley \& Sons (New York). 

 \no [7]  Leibovitz, C., Peres, A.  1963, 
    Annals of Physics,  25,  400.   

 \no [8]  Rindler, W. 1966,  {\it Special Relativity}, Second Edition,
Oliver and Boyd, New York.   

 \no [9]   Singal, A. K. 1997, Gen. Rel. Grav., 29, 1371.   
 
\end{document}